\PassOptionsToPackage{table,xcdraw}{xcolor}

\documentclass[screen]{acmart}

\AtBeginDocument{%
  }

\usepackage{multirow}
\usepackage{float}
\copyrightyear{2025}
\acmYear{2025}
\setcopyright{rightsretained}
\acmConference[LAK 2025]{LAK25: The 15th International Learning Analytics and Knowledge Conference}{March 03--07, 2025}{Dublin, Ireland}
\acmBooktitle{LAK25: The 15th International Learning Analytics and Knowledge Conference (LAK 2025), March 03--07, 2025, Dublin, Ireland}
\acmPrice{}
\acmDOI{10.1145/3706468.3706516}
\acmISBN{979-8-4007-0701-8/25/03}
\begin{document}

\title[Combining Large Language Models with Tutoring System Intelligence]{Combining Large Language Models with Tutoring System Intelligence: A Case Study in Caregiver Homework Support}

\author{Devika Venugopalan}
\orcid{0009-0008-0095-4732}
\affiliation{
  \institution{Carnegie Mellon University}
  \streetaddress{5000 Forbes Ave}
  \city{Pittsburgh, PA 15213}
  \country{USA}
}
\email{devikav@cs.cmu.edu}
\author{Ziwen Yan}
\orcid{0009-0009-4823-7174}
\affiliation{
  \institution{Wellesley College}
  \streetaddress{106 Central St}
  \city{Wellesley, MA 02481}
  \country{USA}
}
\email{zy103@wellesley.edu}
\author{Conrad Borchers}
\orcid{0000-0003-3437-8979}
\affiliation{
  \institution{Carnegie Mellon University}
  \streetaddress{5000 Forbes Ave}
  \city{Pittsburgh, PA 15213}
  \country{USA}
}
\email{cborcher@cs.cmu.edu}
\author{Jionghao Lin}
\orcid{0000-0003-3320-3907}
\affiliation{
  \institution{Carnegie Mellon University}
  \streetaddress{5000 Forbes Ave}
  \city{Pittsburgh, PA 15213}
  \country{USA}
}
\email{jionghal@cs.cmu.edu}

\author{Vincent Aleven}
\orcid{0000-0002-1581-6657}
\affiliation{
  \institution{Carnegie Mellon University}
  \streetaddress{5000 Forbes Ave}
  \city{Pittsburgh, PA 15213}
  \country{USA}
}
\email{aleven@cs.cmu.edu}

\renewcommand{\shortauthors}{Venugopalan et al.}

\begin{abstract}
Caregivers (i.e., parents and members of a child's caring community) are underappreciated stakeholders in learning analytics. Although caregiver involvement can enhance student academic outcomes, many obstacles hinder involvement, most notably knowledge gaps with respect to modern school curricula. An emerging topic of interest in learning analytics is hybrid tutoring, which includes instructional and motivational support. Caregivers assert similar roles in homework, yet it is unknown how learning analytics can support them. Our past work with caregivers suggested that conversational support is a promising method of providing caregivers with the guidance needed to effectively support student learning. We developed a system that provides instructional support to caregivers through conversational recommendations generated by a Large Language Model (LLM). Addressing known instructional limitations of LLMs, we use instructional intelligence from tutoring systems while conducting prompt engineering experiments with the open-source Llama 3 LLM. This LLM generated message recommendations for caregivers supporting their child's math practice via chat. Few-shot prompting and combining real-time problem-solving context from tutoring systems with examples of tutoring practices yielded desirable message recommendations. These recommendations were evaluated with ten middle school caregivers, who valued recommendations facilitating content-level support and student metacognition through self-explanation. We contribute insights into how tutoring systems can best be merged with LLMs to support hybrid tutoring settings through conversational assistance, facilitating effective caregiver involvement in tutoring systems.
\end{abstract}



\maketitle

\section{Introduction}

Caregiver (i.e., parents and members of a child's caring community) involvement positively contributes to student outcomes, including academic performance and motivation \cite{hoover1997parents}. However, limited research has studied caregiver involvement in learning analytics applications, such as tutoring systems and student dashboards. While tutored students learn at approximately the same rate per practiced problem-solving step, large-scale analyses found notable variation in prior knowledge~\cite{koedinger2023astonishing}. If learning rates are near-constant and prior knowledge is variable, more practice opportunities could help close knowledge gaps. Motivational human support beyond cognitive tutoring, which caregivers can provide \cite{hoover1997parents}, is needed to help increase opportunities to practice. To do so, our field must study systems and analytics that enhance human support in tutoring system practice; a key idea we investigated was the need for caregivers to receive instructional support, along with guidance on effective tutoring strategies.

One emerging topic of interest in learning analytics is hybrid tutoring, where students are supported by an intelligent system and a human tutor \cite{thomas2023tutor}. Hybrid tutoring provides both instructional and motivational guidance \cite{thomas2023tutor}; this is similar to roles caregivers assert in homework support \cite{hoover1997parents}. However, caregivers often struggle in proving adequate instructional homework support to students \cite{pelemo2022parental}. To address this issue, recent research has investigated novel ways in which caregivers can be involved in tutoring systems \cite{Nguyen2024-dl}. One promising direction suggested by this line of work is providing instructional support to caregivers. This can bridge knowledge gaps that caregivers often report regarding modern math curricula \cite{pelemo2022parental, Nguyen2024-dl}. Other guidance can include suggesting effective tutoring practices for caregivers. However, while prior work suggested that caregivers appreciate conversational recommendations \cite{Nguyen2024-dl}, they also noted a lack of personalization and contextual relevance in pre-generated messages. The current study addresses generation of caregiver message recommendations in a hybrid tutoring context, in real time. 

LLMs have demonstrated great potential to support learning analytics applications and learners, including conversational support for debugging during problem solving \cite{ma2024teach}, contextual reflection triggers in collaborative learning \cite{naik2024generating}, prediction of self-regulated learning \cite{zhang2024using}, and virtual teaching assistants \cite{liuastep}, among others. These advancements promise to improve human learning through conversational tutoring. While promising, a fundamental challenge in the design of these tools is that foundation models are domain-general: they lack expert knowledge on domain-specific instruction. To circumvent this issue, recent research has suggested using instructional material as prompting aids that induce automated tutoring in a domain \cite{schmucker2024ruffle}. Other work has suggested prompting foundation models with pedagogically meaningful guardrails, such as ensuring that answers are withheld, as is practiced in effective human tutoring \cite{pal2024autotutor}.

We study the integration of LLMs into a middle school tutoring system for equation-solving, Lynnette \cite{long2018exactly}, in a hybrid tutoring context, an emerging area of interest in learning analytics \cite{thomas2024improving}. We designed the \textit{Caregiver Conversational Support Tool (CCST)}, which provides personalized, contextually relevant message recommendations through LLMs for caregivers supporting learners in tutoring systems. To our knowledge, this study is the first to evaluate LLM-generated message recommendations based on both contextual information supplied by a tutoring system (e.g., hint use, accuracy, and next viable problem-solving steps) and tutoring principles. The current study presents two angles on developing and evaluating LLM-integrated tutoring systems. First, we describe prompt engineering experiments combining the tutoring system's instructional model, tutoring principles, and LLM instruction. Second, we describe an evaluation study with ten middle school caregivers involved in hybrid tutoring using the CCST, including prototype feedback and perceptions of generated conversational support during live student support. We answer these two primary research questions: \textbf{RQ1:} How can an LLM best generate conversational recommendations for middle school caregivers based on problem-solving context?
\textbf{RQ2:} How do middle school caregivers perceive these conversational recommendations?

\section{Background}

\subsection{Caregiver Support in Learning}
\label{sec: background: caregiversupport}

Caregivers play a key role in their child's academic success \cite{hoover1997parents}. As homework increasingly moves online \cite{magalhaes2020online}, technology-integrated caregiver support offers new involvement opportunities \cite{mrazek2021teenagers}. Yet, research on supporting caregivers with learning analytics is limited, aside from studies on AI acceptance \cite{martens2024algorithmic}. Previous work has focused on indirect support, like notification features \cite{broderick2011increasing}. However, these do not enable  caregivers to provide direct instructional assistance during homework, missing the potential for caregivers to actively enhance academic performance \cite{hoover1997parents}.

Notably, different \textit{types} of caregiver support can have varying impacts on their child's academic performance~\cite{DOCTOROFF2017103}. Given the importance of effective caregiver involvement in homework \cite{hoover1997parents}, there is a research opportunity to explore how learning analytics can support caregivers in enhancing student learning. Many caregivers struggle with homework support due to unfamiliarity with modern curricula or lack of tutoring experience \cite{Kent2022-ik,Nguyen2024-dl}. Providing instructional insights to bridge these gaps is a promising but underexplored approach. Our study addresses this gap by generating chat message recommendations to support caregiver involvement through a tutoring system.

\subsection{Conversational Support and Hybrid Tutoring}
\label{sec: background: conversationalsupport}

Conversational interactions are central to learning and teaching \cite{slavin1987making, nickow2020impressive}. Effective dialog can guide students through problem-solving processes and provide real-time feedback tailored to their individual needs. Recent research has explored how LLMs can enhance tutoring conversations \cite{pal2024autotutor, schmucker2024ruffle, Lin2024-yd}. For example, Khan Academy developed Khanmigo \cite{khanmigo2024khan}, an LLM-powered tutoring system that uses conversational interactions to assist students across various subjects. 

LLMs offer the potential to enhance hybrid tutoring and support human tutors. Lin et al. \cite{Lin2024-yd} utilized prompting and fine-tuning approaches to develop an automated feedback system for tutor training, guiding tutors in delivering effective tutoring strategies during conversations. Despite the potential of LLMs \cite{schmucker2024ruffle, pal2024autotutor}, a significant gap remains between these LLM-based systems and traditional intelligent tutoring systems (ITS), which are designed with instructional models that evaluate learner performance and provide adaptive instruction \cite{long2018exactly}. In contrast, most current LLM applications in education primarily offer feedback based on textual interactions, and rarely incorporate real-time log data to support student learning. This limitation might reduce the ability of LLMs to offer the level of adaptive, data-driven instruction typical for ITS \cite{Huang2021-fv}. Our work bridges this gap by integrating LLM strengths with the adaptive capabilities of ITS.
\subsection{Large Language Models in Learning Analytics Application}
\label{background:llms}

LLMs have become increasingly relevant in learning analytics \cite{yan2024generative}, demonstrating their effectiveness in various educational tasks, such as providing automated feedback to assist student writing \cite{daiassessing}. LLMs can understand natural language in text form (e.g., students' responses to open-ended questions) from the learning process and generate text-based learning support \cite{yan2024generative}. Notably, recent research employed LLMs for enhancing teaching and learning through dialog-based ITS \cite{schmucker2024ruffle}. 
These applications of LLMs in education show the potential to tailor instruction, aiding student comprehension, and aligning with learning analytics goals to enhance educational outcomes \cite{gavsevic2015let}. Despite their potential, LLMs face limitations and open questions in educational technology applications such as ITS. 

 A major concern is their tendency to hallucinate or produce incorrect information \cite{wang2024large}, which can lead to the dissemination of false or misleading content, confusing students or reinforcing misconceptions \cite{nye2023generative}. Another concern is that while LLMs can provide correct answers, they often do so without sound instructional principles, which can obstruct students' deeper learning and comprehension. 
 To address these limitations, recent research has increasingly focused on the Retrieval-Augmented Generation (RAG) approach \cite{lewis2020retrieval}. RAG allows LLMs to retrieve relevant information from external repositories, such as authoritative educational literature, to enhance their responses. The RAG approach can help reduce the likelihood of LLMs generating hallucinated or pedagogically unsound responses by grounding the generated content in credible, context-specific information \cite{han2024improving}. 
 Despite its potential, the application of RAG techniques in the context of tutoring remains underexplored, which is one of the focuses of the present study. %

\section{Computational Methods, Recommendation Design, and Prompt Engineering}

This study aimed to engineer a tool providing LLM-generated conversational recommendations, incorporating content-level support and expert tutoring principles for caregivers helping their student in a tutoring system.

\subsection{Tool and Instructional Context}
\label{sec:methods:tool}
\subsubsection{Overview}
\label{sec:methods:tool:overview}

The CCST aims to guide caregivers as they support their child during ITS practice. We embedded the CCST in the equation-solving Lynnette tutor \cite{long2018exactly}. Like most ITS, Lynnette provides students immediate feedback as they work through multi-step equation solving, including error-specific messages and hints. Students can practice equations with Lynnette alone, as with a typical ITS; however, the CCST also provides the opportunity for caregivers to join their child's practice and assist them through a chat panel. The CCST monitors the interaction and provides conversational support to caregivers in the form of (1) message recommendations and (2) problem-solving path previews. 

One key advancement of the CCST is including intelligence from instructional models into prompting for addressing pedagogical limitations of LLMs \cite{Stamper2024-ne,koedinger2012knowledge}. We used problem-solving context and adaptive instruction from Lynnette to generate message recommendations for caregivers helping their student during math practice. Like many ITS, Lynnette features a domain model for assessment, feedback, hints, and other adaptive instructional support. In the CCST, we leverage the same student action recorders used for tutoring that trace student behaviors, to prompt LLMs. We classify types of recorded student behaviors to sample from evidence-based human tutoring \cite{thomas2023tutor}. For instance, if the ITS graded the last attempt as incorrect, dialog related to reacting to errors would be sampled.

\subsubsection{Caregiver-Student Interaction Design} 
\label{sec:methods:tool:design}
As students solve math problems with the Lynnette ITS, the CCST provides a button to notify their caregiver through SMS. This SMS includes a link for the caregiver to open Lynnette in their browser and join their child. The following descriptions highlight the caregiver's interactions with the tool.

\begin{figure*}
    \centering
    \includegraphics[width=0.8\linewidth]{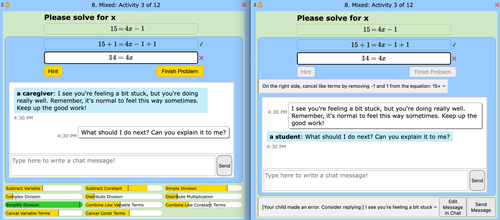}
    \caption{\textbf{Child (left) and Caregiver (right) interface.} The caregiver has a live view of their child's problem-solving steps in Lynnette (top of the screen, both left and right).
    The caregiver also has two dropdown menus, containing: (1) dynamically-generated chat recommendations (bottom) and (2) next-step recommendations based on potential problem-solving pathways (top). }
    \label{fig:userInterface}
\end{figure*}

When the caregiver joins the problem set, a chat panel appears on both the student's and caregiver's screens. The CCST includes the following features for caregivers (see Figure \ref{fig:userInterface}). The first is a real-time display of their child's problem-solving screen. Through synchronization and displaying live problem-solving steps, caregivers can see their child's practice progress. This includes seeing the last submitted graded problem-solving attempt and hint requests. The next feature is instructional guidance through two dropdowns. One dropdown provides suggested next steps in solving the current equation, and is updated as the student submits problem-solving attempts. The second dropdown provides chat message recommendations; caregivers can select and edit these recommendations to send via the chat panel to their child. 
After each problem solving step attempt and chat interaction, the message recommendations dropdown is updated, as an LLM generates new recommendations based on the current problem-solving context (see Section \ref{sec:methods:tool:features}).

\subsubsection{CCST Features and Modules}
\label{sec:methods:tool:features}

\begin{figure*} 
    \centering
    \includegraphics[width=1\linewidth]{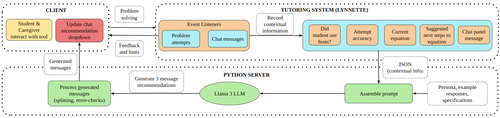}
    \caption{\textbf{CCST Features and Modules.} This figure displays the three main components of the CCST: (1) client, (2) instructional model, and (3) Python server}
    \label{fig:ccstfeatures}
\end{figure*}

The main components of the CCST can be summarized in three modules: the tutoring system's frontend, its instructional model, and LLM Python server, all of which are visualized in Figure \ref{fig:ccstfeatures}. 

The frontend is where caregivers and students interact with each other and the tutoring system. The underlying instructional model of the tutoring system includes student action recorders that trace student and caregiver behaviors for prompt generation and also aid in adaptive instruction. If the student submits an attempt at solving a problem step, or if either the student or caregiver sends a chat message, relevant contextual information is sent from the client to a backend Python server. This includes the current equation being worked on, the last-attempt accuracy, hint usage, and previous chat messages. The tutoring system's instructional model also constructs solution paths to the current equation based on rules outlining viable transformations \cite{long2018exactly}. After classifying whether the solution paths lead closer to the final solution from the current state, up to three next steps, ordered based on their proximity to the desired end state of the problem (i.e., a solution for X) are included in the context sent from the client to the backend Python server.

The real-time problem-solving context described above, along with additional content (e.g., persona, example responses integrating best practices for tutoring, specifications), are assembled into a prompt (for prompt engineering experiments, see Section \ref{sec:prompt-engineering}) that invokes the Llama 3 LLM. This prompt captures problem-solving context and instructs the LLM to provide tutoring advice to the caregiver. The LLM is prompted to generate three message responses in a single run to optimize runtime while also providing caregivers with multiple recommendations. These responses are processed and sent back to the client, appearing in a dropdown on the caregiver interface, as shown in Figure \ref{fig:userInterface}. For this study, we chose to use the open-source Llama 3 8B Instruct model, which is the smallest version of the Llama models. This decision was motivated by our preference to (1) use an open-source model and (2) use a model that could be locally run. The instruct variant is fine-tuned for conversational support \cite{adamsllama3}, and the 8B parameter size was appropriate for running locally. To run the model on our production server, we used Ollama \cite{Liu2024-jn}, an open-source framework to locally run LLMs. Compared to closed-source, proprietary models such as OpenAI's ChatGPT, open-source models bear the advantage of stable and transparent versioning while ensuring equitable access to tool use, irrespective of financial resources and opportunities for comprehensive auditing \cite{casper2024black}.

The tutoring system client and instructional model were hosted using a simple web server running on HTML+JavaScript, which communicated with an additional backend WSGI-compliant Python web server running the LLM. Our server could support the model with 32GB of RAM and 8 Intel Xeon CPUs. As each instance of (a) the caregiver or student sending a message in the chat panel or (b) the student submitting a problem-solving attempt results in a POST request to the server, we further implemented a load balancer that only allows a request if more than 30 seconds has elapsed since the last request. This timeframe was determined based on estimated average prompt processing times in the LLM module. The client sends the aforementioned contextual information in JSON format. This is an example payload of a request resulting from a sample interaction: $\{$``chat message'': ``Caregiver: hey what do you need'', ``next step'': [``Subtract 1 from both sides: x-1+1 = 3-1''], ``used hint'': ``False'', ``accuracy'': ``error'', ``question'': ``1+x=3''$\}$.

\subsection{Prompt Engineering Iterations}
\label{sec:prompt-engineering}
The CCST's caregiver chat message recommendations are generated by invoking the Llama 3 LLM with a prompt integrating contextual information.
We defined the following properties as desirable for these messages.
First, following tutoring best practices, such as assessing a student's prior knowledge, responding to errors in a way that increases motivation and engagement, and giving effective praise that acknowledges effort \cite{thomas2023tutor}. 
Second, including brief, explanatory text at the beginning of each message (formatted as [explanation]: message), where [explanation] highlights the main goal of the message. Examples of these explanations are ``Ask to self-explain,'' and ``Praise your child for a correct response.'' Including these explanations at the front of messages allows caregivers to understand the objective of each message recommendation \cite{Nguyen2024-dl}.
Third, being contextually relevant to the live tutoring taking place. A key novelty of our approach is integrating user-specific problem-solving context, along with principles of effective tutoring, while generating message recommendations. Rather than having generalized messages, the LLM-generated messages are personalized to the current interaction, taking into account factors like the equation being worked on, hint usage, and the student's accuracy. These are reflected in generated responses, as shown in Table \ref{table:responses}. 

We evaluated our prompt engineering experiments using the CLEAR Framework for Prompt Engineering \cite{LO2023102720}, a method to optimize interactions with LLMs. The five core principles of this framework involve determining whether a prompt is concise, logical (structured and coherent), explicit (clear output specifications), adaptive (customizable), and reflective (continuous evaluation and improvement of prompt). We conducted seven main prompt engineering rounds, iterating upon previous rounds until the quality of generated responses was satisfactory.\footnote{The seven main prompts are included in this GitHub repository, along with example code for our server architecture involving the use of different contextual information and assembling them into a prompt: https://github.com/devika-prog/Caregiver-Conversational-Support-Tool} 

We grouped the prompts used in the seven rounds into three categories, each which served as responses to shortcomings of the previous category. The CLEAR framework does not prescribe specific sample sizes, so we chose to evaluate about 50 to 80 examples of generated responses per prompt experiment.

The prompts in \textbf{Category 1} (prompts 1 - 3) followed zero-shot prompting, in which no examples of desired output were provided. These prompts provide information about the task and formatting guidelines \cite{White2023-bi, Lin2024-yd}. Limited contextual information (only previously sent chat panel messages) was provided. Prompt 1 instructed the LLM to assume the appropriate persona \cite{White2023-bi} and included a list of all previously sent chat panel messages. Prompt 2 specifically stated the purpose of using the list of chat panel messages (to generate a message that a caregiver would say to their child at this point in the conversation). Prompt 3 provided guidelines for length and explicitly noted that chat panel messages were delimited \cite{Lin2024-yd}, allowing the model to differentiate between them and the instructions in the prompt.

The prompts in \textbf{Category 2} (prompts 4 and 5) used few shot prompting, in which examples of desirable responses were provided. These prompts incorporated techniques from recent advancements in prompt engineering and were primarily motivated by RAG, which involves the use of content from external sources to enhance an LLM's capabilities \cite{han2024improving}. We integrated knowledge of tutoring principles \cite{thomas2023tutor} by providing literature-supported example responses in a section of the prompt.  Tutorial dialog is most effective when it elicits opportunities to learn and reflect, based on the theory of ``accountable talk'' by Resnick et al. \cite{resnick2018accountable}. Such opportunities arise when dialog allows students to reflect, explain, and challenge their position. Accordingly, our integrated tutoring practices include responding to student errors, assessing what a student knows, and providing effective praise, all known to increase self-efficacy and engagement \cite{thomas2023tutor}. Prompt 4 provided seven example tutoring responses, each of which was preceded by detailed explanations of why those responses are characteristic of effective tutoring. Prompt 5 removed these lengthy explanations of tutoring strategies, and instead separated responses into three general categories: responding to errors, determining what a student already knows, and giving praise. This prompt included 3 examples per category of response.

The prompts in \textbf{Category 3 }(prompts 6 and 7) followed our novel approach of incorporating instructional intelligence received from the ITS. These prompts continued to use few-shot prompting as the prompts in Category 2 did, but included a new section integrating session-specific problem-solving context. This content included (1) whether the student's problem-solving attempt was correct or not, (2) whether the student used hints, (3) a list of all previously sent chat panel messages, (4) the current equation needing to be solved, and (5) suggested next steps to solving the equation. Notably, these suggested next steps were generated using the tutoring system's underlying instructional model, as described in Section \ref{sec:methods:tool:features}. Prompt 7 was structurally the same as Prompt 6, but added examples of including short explanatory messages to the front of each generated response \cite{Nguyen2024-dl}. This feature aids caregivers in deciding which of the messages to send, by briefly summarizing the objective of each message.

\begin{figure*} 
    \centering
    \includegraphics[width=1\linewidth]{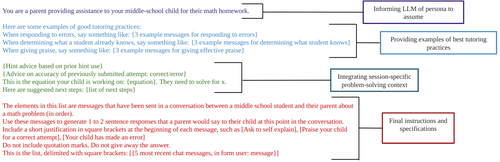}
    \caption{\textbf{Prompt 7 Breakdown.} Prompt components with their purpose. The section integrating session-specific problem-solving context is unique to each interaction, and its different components are represented by the placeholders in the figure. This is an example of what that section of the prompt could include: ``Your child did use a hint, so ask them what they understood from the hint. Your child made an error, so you should focus on responding to the error as described earlier. Here are suggested next steps: Subtract 5 from both sides: 15-5 = -2x-5+5. This is the equation your child is working on: 15 = -2x-5. They need to solve for x.'' }
    \label{fig:prompt7}
\end{figure*}

Figure \ref{fig:prompt7} outlines the structure of Prompt 7. 
The following are 3 examples of recommendations that the LLM generated in response to Prompt 7: \textbf{(1)} \textit{$[$Walk through hints$]$ I love how you tried very hard and focused on the problem! Can you tell me what you understood from the hint about dividing both sides by the coefficient of x?} \textbf{(2)}\textit{$[$Ask to self-explain$]$ Can you walk me through your thinking in this step? Why do you think we should divide both sides by 6?} \textbf{(3) }  \textit{$[$Praise your child for a correct attempt$]$ Great job on getting this far! It looks like you’re close to solving it. Can you tell me what you did first?}

After developing each prompt, we evaluated prompt quality with the CLEAR framework \cite{LO2023102720} before using the prompt to invoke the LLM. Our prompts were (1) concise, providing solely the information needed, (2) logical, with a clear structure of components, as shown in Figure \ref{fig:prompt7}, (3) explicit, stating criteria for messages, including parameters like length and content, (4) adaptive, as illustrated by the integration of real-time problem-solving context, and (5) reflective, as shown by our multiple iterations of prompt engineering to address shortcomings of responses from previous prompts.

\subsection{Prompt Engineering Results}
\subsubsection{Generated Response Characteristics}

In the following section, we describe the characteristics of generated responses when invoking the LLM with prompts from each of the 3 main categories described in Section \ref{sec:prompt-engineering}. Each response was evaluated against the `desirable' standards established in Section \ref{sec:prompt-engineering}.

\textbf{Category 1}: Chat panel messages were recognized, and content from the messages was integrated into the LLM's generated recommendations.  It appeared, however, that the LLM would benefit from additional contextual information in the prompt (e.g., the problem being worked on) since the generated responses often seemed to lack context. Some responses also had phrasing that would not be used in normal conversation in this context, as highlighted in this example: \textit{ ``Ah okay sweetie! I’m here to help. Take a deep breath and let’s take a look at the problem together. Which one is giving you trouble? Is it a multiplication or division question? Or maybe it’s something with decimals or fractions? Let me see what you’re working on and we’ll figure it out together!"}

\textbf{Category 2}: Chat recommendations integrated best practices for tutoring and closely resembled provided examples. Since little problem-solving context was provided, the recommendations were still not specific to the given problem-solving context, as shown in this example \textit{``Let's try solving the problem together. Can you tell me what you did first?''}

\textbf{Category 3}: The recommendations were more specific to the current interaction, appropriately integrated problem-solving context, incorporated best tutoring practices, and included explanatory messages to categorize different types of responses, as evidenced in this example: \textit{``[Ask to self-explain ] Can you walk me through your thinking in this step? Why do you think we should divide
both sides by 6?''} Responses generated by Category 3 prompts were evaluated to be the most desirable according to our standards, mentioned in Section \ref{sec:prompt-engineering}.

\subsubsection{Incorporation of Contextual Information}
\label{sec:prompt-engineering:contextualinfo}
An objective of this study was to determine how integrating real-time problem-solving context into a prompt allows the LLM to generate personalized, contextually relevant recommendations. Table \ref{table:responses} demonstrates how different values for the 5 problem-solving context variables (see Section \ref{sec:prompt-engineering}) resulted in different content in generated recommendations. The examples included in the table are from simulated trials where a research team member provided different inputs to the tutoring system (e.g., submitting correct and erroneous responses, sending chat panel messages, using hints) to determine how generated recommendations were affected.

\begin{table*}
\footnotesize
\caption{\textbf{Integration of Problem-Solving Context.} Demonstrates how different values for problem-solving context variables modified characteristics of generated recommendations.}
\begin{tabular} { |p{1.25cm}||p{3cm}|p{3cm}|p{3cm}| p{3cm} |  } 
 \hline
 \multicolumn{5}{|c|}{\textbf{Problem Solving Context and Generated Recommendations}} \\
 \hline
 \textbf{} & \textbf{Example 1} & \textbf{Example 2} & \textbf{Example 3} & \textbf{Example 4} \\
 \hline
 \textbf{Current Equation} &  6x = 12 & 6x = 12 & 6x = 12 & 6x = 12 \\

 \hline
  \textbf{Suggested Next Steps} &  Divide both sides by the coefficient of x, which is 6:    
  6x/6 = 12/6 &  Divide both sides by the coefficient of x, which is 6: 6x/6 = 12/6 &  Divide both sides by the coefficient of x, which is 6: 6x/6 = 12/6 &  Divide both sides by the coefficient of x, which is 6: 6x/6 = 12/6 \\
 \hline
 \textbf{Used Hints} &  True  & False & True & False \\
 \hline

 \textbf{Attempt Accuracy} & Error & Error & Error & Correct \\

 \hline
 \textbf {Chat Panel Message} &  Student: ``I need help"  & Student: ``I need help"  & Student: ``Should I divide by 6?"  & Student: ``I think I got it" \\
 \hline

  \textbf{Generated Response (Output)} &  [Walk through hints] Can you tell me what you understood from the hint about dividing both sides by the coefficient of x? & [Request a hint] How
about you request a hint and walk me through it? & [Ask to self-explain]
Can you walk me
through your thinking
in this step? Why do
you think we should
divide both sides by 6? & [Praise your child for a
correct attempt] Great
job on getting this far!
I appreciate how hard
you worked on this
problem.\\
 \hline
   \textbf{Evidence of Instructional Intelligence} & Student did use hints and made an error, so recommendation prompts student to explain hint, to encourage further thought about the problem. & Student has not used
hints and made an error, so recommendation is to request a
hint to assist with their problem-solving. & Student sent a chat
message about dividing by 6, and since
they made an error,
recommendation asks them to explain their thought process. & Student’s attempt accuracy was correct,
so recommendation is
praising student in a way that follows best tutoring practices. \\
 \hline
  
\end{tabular}
\label{table:responses}
\end{table*}

\section{Prototyping Study Methods}

The objective of this case study was to explore how LLM-generated conversational recommendations can support caregivers during synchronous collaboration with their child, as their child practices with a tutoring system. Caregivers, participating alongside their children in dyads, were recruited through a summer program for middle school students at a university in the Northeastern United States. A semi-structured 60-minute prototyping interview was conducted during the second day of a two-day design workshop. The same protocol was also conducted with caregiver and student dyads who opted to participate online via Zoom. This study focuses solely on caregiver perspectives, although the workshop protocol also included activities and perspectives of students.

\subsection{Sample}
\subsubsection{Recruitment and Compensation}
Participants for the two workshops were recruited through an outreach center affiliated with a university in the Northeastern United States. IRB approval for this study was obtained, and informed consent was taken from human subjects. For the remote interviews, caregiver-student dyads were recruited using contacts from prior design interviews in the project, primarily through social media and email lists from local caregiver outreach centers. Remote interviews were conducted after the workshops to reach a satisfactory sample size, determined by reaching saturation of qualitative insights related to our research questions, as is common practice in qualitative research \cite{saunders2018saturation}. Participants in the workshops received a \$70 Amazon gift card each upon completion of both 2-hour sessions. For remote interviews, each caregiver and student received a \$25 Amazon gift card as compensation.

\subsubsection{Participation}
A total of ten caregivers participated in the study, with four attending the workshops and six attending remote interviews. The participants were 40\% Asian and 60\% White, comprising 20\% male and 80\% female, with an average age of $M$ = 47.9 years ($SD$ = 6.3). The language of participation for all caregivers was primarily English, though one participant with limited English proficiency responded to some questions in Chinese.

\subsection{Material and Procedure}

\subsubsection{Workshop}
The workshop aimed to gather suggestions for improving the design of the CCST based on caregiver needs and explore their perceptions of LLM-generated conversational support in caregiver-student interactions (RQ2).%

Participants were divided into groups of one to three depending on the number of participants and facilitators, with each group guided by a designated facilitator, who recorded conversations using audio recorders. To simulate a remote homework support session, caregivers and students were seated separately and communicated through the chat panel. During this time, they worked on problem sets in the ITS together, with caregivers instructed to send chat messages to the student, check the hint function, and provide comments on these features. The facilitator guided these interactions and asked caregivers questions about the length, tone, quantity, and response time of the messages, as well as typical homework routines, opinions on tool functionality, and conversational support. 
To obtain more in-depth feedback, we provided a printout of additional message recommendations, which were based on examples generated through prior testing of the CCST as well as best tutoring practices \cite{thomas2023tutor}. Caregivers also annotated screenshots of the interface, which show a sample interaction, to suggest design improvements \cite{Solyst2023}. 
While the tool was fully implemented during the session, these inquiries and additional materials, along with discussions on design feedback and various use case scenarios, provided a deeper understanding of caregiver preferences.

\subsubsection{Remote Interview}
Each interview session lasted for one hour, and followed the same procedure as the in-person protocol. The student and caregiver were seated in separate rooms and interacted with the tool independently to simulate remote use. Screenshot annotations were omitted due to the challenges of printing and sharing annotations via Zoom. Rather, we asked participants to verbally describe any changes they would like to see and how they would implement them, allowing us to gain insights into the design solutions they envisioned. The sessions were recorded using Zoom's built-in recording feature, capturing both the breakout rooms and the main room.

\subsection{Data and Processing}
Following best practices from prior work \cite{zhang2024using}, all audio records of participants’ sessions were transcribed using OpenAI Whisper \cite{pmlr-v202-radford23a} or Zoom's built-in transcription tool. Whisper transcribed and translated utterances in Mandarin Chinese from one participant with limited English proficiency. A research team member fluent in both Mandarin and English reviewed the translations and confirmed that they were of sufficient quality for thematic analysis. Another team member recorded the handwritten notes and drawings from the annotated screenshots into a spreadsheet for further analysis.

\subsection{Data Analysis Methods}
The qualitative data (i.e., interview and annotation data) was analyzed using a thematic analysis approach with an open coding scheme. Two research team members independently conducted a first round of inductive open coding to establish initial descriptive codes \cite{wolcott1994transforming}. 
A final round of discussion and consolidation of the resulting topic centers was held to eliminate individual coder bias \cite{Szymanski2024}.

\section{Prototyping Study Results}

The following sections address RQ2 by presenting themes describing caregiver preferences for conversational support. 

\textbf{\textit{Theme 1: Caregiver preference for content-level support over motivational support.}} 

Caregivers demonstrated a preference for recommendations providing mathematical guidance rather than motivational support. Six caregivers preferred more direct, content-focused recommendations, two caregivers found both types of support useful, and one caregiver preferred to provide motivational help independently of the system. We identified two subthemes that further highlight the underlying reasons for this preference. \textbf{\textit{Caregivers find content-level messages valuable for providing the support they often feel unequipped to offer}}: Messages providing content-level guidance are especially beneficial for caregivers who may feel less confident in their ability to assist with math homework. Three caregivers highlighted how these recommendations help guide them through complex problems, providing the support needed when they are unsure of the correct answers or fear giving incorrect guidance. As Caregiver C10 noted, ``I think it’s nice to have these options [recommended messages] because sometimes you get frustrated when you don't know the right things to say . . . you might say the wrong thing.'' \textbf{\textit{Caregivers did not find motivational support as helpful due to its misalignment with their current tutoring practices}}: Caregivers who favored content-level support typically adopt a direct, results-oriented approach when tutoring their children, leading them to find motivational support as not useful. Caregiver C6 described motivational support as ``just filler'' and ``not as useful.'' Other caregivers noted that their children ``wouldn't probably react well to praising their effort'' (C10) or ``just want to solve the problem'' (C4). 

\textbf{\textit{Theme 2: Caregivers preferred messages that prompt student metacognition (i.e. explanation of thought processes), as it provided them with deeper insights into their child's thinking.}} 
While the CCST generally allowed for live synchronization of student problem-solving steps, allowing the caregiver to see a reflection of their child's screen (see Section \ref{sec:methods:tool:design}), caregivers articulated a desire for more in-depth insight into their child's thinking. Five caregivers preferred messages that prompt students to explain their reasoning process more thoroughly. For example, C6 mentioned asking ``tell me your thought process'' and ``how you're thinking about this.'' C4 emphasized the importance of getting more information by looking at the ``work paper'' of the student, which may refer to scratchpad notes students craft before entering problem-solving step attempts (visible to caregivers) into the tutoring system. Similarly, C8 suggested asking ``why does the work help you solve the equation.'' 
Caregivers noted that understanding student thought processes can allow them to measure knowledge gaps, which in turn allow for more targeted support. As C9 noted, ``Walk me through it and can you explain to me what you did here. I think those are good they reinforce that the need to be able not just to come up with an answer but to explain and to show your work in whatever manner is required.'' Another caregiver C8 added that ``I would say ask them just to speak out their thought process loud, so I know where the mistake is. So, you know, it can be a careless error right and or they just don't know how to do it. So I want to find out why they did it wrong.'' 

\subsection{Technical Feedback and Feasibility of Generated Messages}
\label{sec:technical_feedback}

Caregivers also provided feedback covering technical aspects of the messages (i.e. length, style), as well as the effectiveness of incorporating problem-solving context. 
\textbf{\textit{Contextual Information:}} The inclusion of problem-solving context (see Table \ref{table:responses}), specifically, incorporating whether the student used hints or made errors, was recognized and appreciated, since it provided additional opportunities to engage with and understand instructional content. C8 emphasized that ``They [messages] should be based on the mistakes or the steps of which [student] got wrong to have customized.'' C5 expressed ``Oh that's good, `How about you request a hint and walk me through it?' Cuz he may have just asked me before he hit the hint.'' This context allows caregivers to gauge the student's current understanding and guide them according to their current progress.
\textbf{\textit{Message Length:}} Six out of nine caregivers found the messages to be too long, making them hard to process during live tutoring. For example, C6 noted that: ``Like having to read through this text is cumbersome.'' \textbf{\textit{Number of Messages:}} While eight out of nine caregivers felt the number of messages (i.e., three recommendations at a time) was adequate, one parent preferred a narrowing down to only one message. \textbf{\textit{Message Style:}} Caregiver feedback on tone was mixed. Three out of seven caregivers found the tone of the message recommendations lacking, each citing that the tone is inauthentic and artificial. C5 noted: ``Glad you’re focused just seems artificial to me.'' Conversely, C10 valued the gentle tone, stating, ``It's not too harsh when it says you might have made a mistake, like, using the word `might' or `small'. I think the tone was good.'' The remaining four caregivers found the tone positive and adequate, highlighting that it is encouraging. C8 remarked, ``I think the tone is good. I mean, it's very encouraging. It's very positive.'' \textbf{\textit{Latency:}} Two out of six caregivers felt that the message generation was somewhat slow, while four found it to be reasonably quick. C9 noted that ``it's faster than us,'' highlighting that the performance is generally satisfactory, but could potentially be improved.

\section{Discussion}

The objective of the current study is to determine how to effectively generate conversational support for caregivers providing guidance as their child practices with an ITS. We did so by combining the instructional model of a tutoring system and the language abilities of an LLM, aiming to address LLM limitations for instruction identified in prior research \cite{Stamper2024-ne}. We then determined caregiver perceptions on LLM-generated conversational support by conducting interviews in which we tested the Caregiver Conversational Support Tool (CCST) with middle-school caregivers.
\subsection{Conversational Support Message Generation}

RQ1 focused on how tutoring system capabilities and best tutoring practices can be provided to an LLM to generate conversational recommendations. We conducted prompt engineering experiments integrating tutoring system context data (e.g., correctness, problem-solving pathways) with LLM instructions using the open-source Llama 3 LLM. 

Recent research has argued that LLMs lack instructional principles, such as determining a student's prior knowledge, to suffice as tutoring agents \cite{Stamper2024-ne}. We found that the LLM was able to adequately integrate problem-solving context into generated message recommendations. We \textit{also} provided the LLM with guidance as to \textit{how} it could use that data from a tutoring system to provide suggestions in generated responses (see Figure \ref{fig:prompt7}). We observed that such instructions, alongside few shot examples on effective tutorial dialog \cite{thomas2023tutor}, were especially useful for generating message recommendations attuned to student actions. Further, by providing specific solution options as to how the student may solve the equation, we observed no issues related to providing incorrect, hallucinated math advice. While the tutoring system can generate correct solutions without error due to its instructional model, an LLM alone lacks arithmetic abilities \cite{han2024improving,zhang20243dg}. We only observed such hallucinations when limited tutoring system context was given in Category 1 prompts (see Section \ref{sec:prompt-engineering}). Taken together, we observed that generated messages were not only pedagogically fit for tutorial dialog \cite{thomas2023tutor}, but also more reliable in terms of arithmetic and tutoring advice. 

\subsection{Prototyping Study Discussion}

RQ2 focused on caregiver perceptions of conversational recommendations integrating tutoring system and LLM intelligence. Design research employing the CCST with ten middle school caregivers provided tangible insights into how caregivers can be best supported in homework support, which is crucial for student learning \cite{hoover1997parents,Nguyen2024-dl}.

\textit{\textbf{Caregivers appreciated content-level support and desired further opportunities to engage with content:}} Caregivers appreciated both content-level and motivational support messages during tutoring but saw greater value in content-level support. They appreciated messages that integrated instructional guidance in messages, including the ``Suggested Next Steps,'' which provided concrete instructions as to how to solve an equation. These steps were particularly helpful for caregivers who felt less confident in providing accurate math help, as they offered clarity and reassurance, which aligns with prior research on caregiver support needs \cite{Nguyen2024-dl,pelemo2022parental}. Caregivers also valued the integration of real-time problem-solving context (i.e., student hint use and errors) within content-focused messages, especially when it helped them and their child to engage with the content. Future designs should aim to integrate such data-driven contextual information while ensuring the student-specific guidance remains actionable. This involves enhancing message specificity to align with student actions and potentially avoiding the repetition of information available in the tutoring system. Some caregivers also noted that motivational messages were less aligned with their homework support style. Future design revisions could allow caregivers to customize which messages to receive.

\textbf{\textit{Conversational recommendations can help caregivers to understand student metacognition:}} We found that caregivers favored messages that encouraged self-explanation, an effective instructional strategy \cite{aleven2002effective} that helps students articulate their reasoning and reflect on their problem-solving~\cite{renkl1999learning}. %
Future improvements should focus on tailoring these self-explanation prompts to better align with caregiver preferences. Specifically, caregiver feedback underscores the need for personalized tone in messages, and potentially improved relevance and timeliness. One major challenge identified is balancing the specificity of context with the timeliness of responses, given delays in message generation. Future designs may adapt recommendations to caregiver voice and investigate when the generative capabilities of LLMs, as opposed to pre-generated messages, may be most beneficial for conversational support in tutoring.

\subsection{Implications}

Caregivers describe lack of instructional support as a key barrier to getting involved with their child's online homework \cite{Nguyen2024-dl}. Hence, one central goal of the CCST was to aid caregivers in bridging knowledge gaps while supporting their child during tutoring system practice. Our results suggest that this instructional support could help caregivers in providing instructional interventions in their child's math homework \cite{hoover1997parents}. The caregivers we interviewed in this study especially favored message recommendations that targeted content-level support, which were informed by tutoring system instruction and log data. Hence, as a broader implication to the field, content-level conversational support through LLMs may augment increasingly common hybrid tutoring scenarios to improve student learning \cite{thomas2023tutor}. Moreover, addressing recently identified instructional limitations of LLMs \cite{Stamper2024-ne}, our findings imply that tutoring system instruction and log data collected during practice can be productively integrated into LLMs to help improve the instructional rigor of LLM generations in educational contexts similar to the one studied here.

For productive merging of tutoring systems and LLMs, we note that providing tutoring system data into prompting is not enough; it requires instructions and examples as to \textit{how} these data should be used. Thus, other applications providing context to better inform foundation models should accompany data with \textit{explanation} when prompting an LLM, for example, by describing tutoring best practices, as done in this study \cite{thomas2023tutor}. This method of using an LLM and ITS in conjunction is a key contribution to the present study.

Our designs and methods may be transferred to other systems supporting learners that are common in learning analytics. For instance, collaborative learning systems, where students work together to solve problems, might benefit from the incorporation of an LLM providing instructional support. It can be challenging for middle-school students to effectively collaborate, even when rule-based support that interprets conversations is present \cite{borchers2024combining}, so integrating an LLM that provides suggestions based on instructional context from tutoring systems could guide student collaboration. Our findings may also inform systems involving no live human interaction. For instance, the JeepyTA \cite{liuastep} uses an LLM as a virtual teaching assistant. Although JeepyTA's initial evaluation found it effective in providing information about a course, it lacked capabilities to sustain student motivation and contribute to effective learning, which might be addressed through responsive and motivational message support reported here. Overall, integrating effective tutoring practices and tutoring system intelligence into LLMs may enhance personalized support in online learning. However design research with end-users will be important for sustainable adoption and improved learning. There is also a potential to expand the scope of our tool. Our current focus is math, but could be broadened to other subjects, as tutoring systems span several STEM domains such as physics, chemistry, and logic \cite{aleven2002effective}.

\subsection{Limitations and Future work}
We acknowledge three study limitations. First, our hardware and resource limitations of this project limited the speed of message generations when running LLMs, which could be further improved through GPUs. Average latencies for message recommendations in this study were about 25 s, which at times was perceived as too slow by caregivers, especially when reference was given to specific actions in the tutoring system. Future work may investigate how recommendations can better deal with latency, for example, by updating recommendations to include less specific advice if the tutoring system detected the student to have moved on from when the request was first sent. Another point of exploration is using different LLMs beyond Llama 3.

Second, our prompt engineering method was purely qualitative, using the CLEAR framework \cite{LO2023102720}. While this method is suited for quick prompt iterations, future research may study the emerging characteristics of message recommendations through sentence embeddings, including but not limited to, clustering, to refine prompting. Quantitative curation of message recommendation may further be used to fine-tune LLMs toward more desirable output.

Third, participants may not necessarily be representative of the larger caregiver population. Considering that participants were volunteers, there is a potential bias that these caregivers are especially engaged in homework support. In addition, most were women, and all were either White or Asian. To ensure equity and inclusivity of our tool's use and design, future work could involve recruiting a larger and more diverse sample for design research.
\section{Summary and Conclusions}
This study investigated integrating tutoring systems and LLMs to improve conversational support in hybrid tutoring, where caregivers and tutoring systems assist students in problem solving. While previous research highlights the value of caregiver involvement in homework, little learning analytics research has studied how caregivers can be supported in asserting practice support roles. Our results show that using instruction and log data from tutoring systems in prompts enables LLMs to generate contextually relevant messages for hybrid tutors. The LLM adapted its responses based on problem-solving context, and we found that effective prompting requires not just data, but also example responses. Based on design research, caregivers valued content-level message recommendations from our system and especially appreciated those prompting their child to explain their thought process. Such metacognitive prompts show promise in helping caregivers support students in their homework constructively, which not all caregivers do or are able to. A key contribution of this work is demonstrating how LLMs and tutoring systems can work together during problem solving, using real-world data to provide caregivers with effective tutoring guidance. This approach enhances personalized, contextually relevant conversational support, advancing learning with intelligent tutoring systems.

\begin{acks}
This research was funded by the Institute of Education Sciences (IES) of the U.S. Department of Education (Award \#R305A220386). D. Venugopalan was supported by the Summer Undergraduate Research Apprenticeship (SURA) program at Carnegie Mellon University. Z. Yan was supported by the National Science Foundation (NSF) Research Experiences for Undergraduates (REU) program.
\end{acks}

\bibliographystyle{ACM-Reference-Format}
\bibliography{main}

\end{document}